\documentclass[conference, 10pt]{IEEEtran}
\IEEEoverridecommandlockouts

\usepackage[
  left=19.1mm,
  right=19.1mm,
  bottom=22mm,
  top=19.1mm,
  headheight=0pt,
  headsep=0pt
]{geometry}

\makeatletter

\usepackage{graphicx} 
\usepackage{lipsum}
\usepackage{amsmath}
\usepackage{amsfonts}
\usepackage{amssymb}
\usepackage{bm}
\usepackage{tikz}
\usepackage{comment}

\usepackage{amsthm}
\usepackage{xintexpr}
\usepackage{xinttools} 
\usepackage{tikz}
\usepackage{xcolor}
\usepackage{hyperref}
\usepackage{cleveref}
\usepackage{pgfplots}
\pgfplotsset{compat=1.18}
\pgfplotsset{colormap={violet}{rgb255=(25,25,122) rgb255=(238,140,238) color=(white)}}

\usepackage{subcaption}

\newcommand{\R}{\mathbb{R}}

\newtheorem*{remark}{Remark}

\usepackage{cite}
\def\BibTeX{{\rm B\kern-.05em{\sc i\kern-.025em b}\kern-.08em
    T\kern-.1667em\lower.7ex\hbox{E}\kern-.125emX}}

\crefname{figure}{Fig.}{Figs.}
\Crefname{figure}{Fig.}{Figs.}

\begin{document}

\flushbottom

\title{Closed Form Modelling and Identification of Banking Effects in Confined Waters
\thanks{The research in this paper is funded by Orients Fond Denmark}}

\author{
    \IEEEauthorblockN{Jeppe H. Mikkelsen\IEEEauthorrefmark{1}, Thomas T. Enevoldsen\IEEEauthorrefmark{2}, Bugge T. Jensen\IEEEauthorrefmark{3}, \\ Michael Jeppesen\IEEEauthorrefmark{4}, Roberto Galeazzi\IEEEauthorrefmark{1}, and Dimitrios Papageorgiou\IEEEauthorrefmark{1},}
    
    \IEEEauthorblockA{\IEEEauthorrefmark{1} Technical University of Denmark, Institute of Photonics and Electrical Engineering, \{jehmi, roga, dimpa\}@dtu.dk}
    
    \IEEEauthorblockA{\IEEEauthorrefmark{2} ABB A/S, Marine and Ports, thomas.t.enevoldsen@dk.abb.com}

    \IEEEauthorblockA{\IEEEauthorrefmark{3} Vessel Performance Solutions, btj@vpsolutions.dk}

    \IEEEauthorblockA{\IEEEauthorrefmark{4} Force Technology, mcj@forcetechnology.com}
}

\maketitle

\begin{abstract}
    Vessels navigating in confined waters are subject to banking effects, which are hydrodynamic forces and moments arising from pressure differentials between the vessel sides, significantly affecting manoeuvrability and safety. Existing numerical approaches such as computational fluid dynamics (CFD) can accurately capture these effects but are computationally expensive and unsuitable for real-time control or estimation. This paper presents a closed-form, first-principles model of banking effects. The model coefficients are identified using physics-informed regression on towing tank experiment data for a scaled container vessel. Validation through Shapley value analysis confirms the significance of the banking terms in reproducing the measured forces and moments. Lastly, the derived coefficients are shown to be non-dimensional, making the model applicable across different scales that preserve vessel geometry.
\end{abstract}

\section{Introduction}
Vessels operating in confined waterways are subject to hydrodynamic forces and moments collectively known as banking effects, which can significantly impact manoeuvrability. These effects can challenge navigational control, particularly in narrow channels, and in extreme cases, may lead to vessel grounding. A prominent example is the 2021 \textit{Ever Given} incident, where a large container ship became lodged in the Suez Canal, halting traffic for six days and blocking a vital route for international shipping. Given the severity of such incidents, there is a clear motivation to develop models capable of capturing bank effects for use in warning systems, automated navigation, and control applications.


Bank effects arise due to pressure differences between the port and starboard side of the vessel, induced by the flow of water along the hull of the vessel. These pressure differences can be calculated based on numerical methods such as computational fluid dynamics (CFD). In \cite{Liu2020CFD-BasedShip} the authors present a CFD study on the effect of varying water depth, forward speed, and rudder angle on banking effects. Their main finding showed that as the water depth decreases, the banking effects turn from attractive to repulsive. Furthermore, they performed analysis on the force distribution on the hull of the ship, showing that the bow and stern are mainly subject to repulsive forces and the midship is mainly subject to attractive forces. They also found that the repulsive forces at the bow of the ship dominates the repulsive forces at the stern, producing a bow-away yaw moment. In \cite{Liu2021CFDConditions}, they further validate their simulation results using experiment data. In \cite{Oud2022CFDShips}, the authors investigate shallow water effects on ships, using CFD analysis and towing tank experiments. They found that the hull forces increase with decreasing water depth, that the sinkage and trim due to squatting increases with decreasing water depth and increasing drift angle, and that low speed tests can lead to discrepancies between CFD and experiments due to the presence of both laminar and turbulent flow, which cannot be captured using CFD. In \cite{vanHoydonck2015BankKVLCC2}, towing tank experiments were used to validate the use of CFD analysis for banking effects. It was found that the distance to the bank has a major effect on the bank suction, but a less pronounced effect on the bank cushioning. Furthermore, they found that with a decreasing under keel clearance, the pressure along the underside of the ship decreases leading to squatting. Furthermore, a decrease in the under keel clearance also leads to more water passing along the sides of the ship, exacerbating the bank suction. Lastly, the authors found that the addition of a propeller induces a pressure decrease at the aft of the ship, leading to a yaw moment pulling the stern towards the bank. As an alternative to CFD analysis, in \cite{Huang2020AWater} a potential flow method is presented for calculating banking and squatting forces for a vessel moving in a canal. In \cite{Degrieck2021HydrodynamicStudy}, a comparative study between potential flow and CFD was performed. It was found that potential flow methods are incapable of adequately capturing banking effects due to the presence of waves and viscous effects. While CFD methods can provide banking effect forces, they are computationally intensive and typically unsuitable for real-time applications. Furthermore, they are also not amenable to control applications, since they use numerical methods to calculate banking effects. In order to use banking models for real-time applications in estimation, planning, and control, the banking models need to have a closed form. An early proposed polynomial model for banking effects relying on a bank-distance parameter, Froude number, draught-to-underkeel-clearance ratio, bank slope, and propeller thrust was presented in \cite{ChNg1993AWater}. In \cite{Vantorre2003ExperimentalForces}, the authors proposed a first principles model relying on Bernoulli's principle, also accounting for the additional banking induced by the ship propellers. In \cite{Yasukawa2019ManeuveringBank}, the authors present a closed form model based on an extended first principles model using Bernoulli's principle, with the addition of a polynomial term for capturing the effect of the vessel states on the banking effect. In \cite{Mai2023AShip} the authors use CFD simulations to generate captive tests used for estimating the coefficients of a closed form banking model.

Although these methods are based on Bernoulli's principle, they mainly rely on polynomial coefficients which are difficult to interpret. The main contribution of this paper pertains to the modelling of banking effects using first principles in order to arrive at an explainable model that can be used in estimation, control, and planning applications. Furthermore, the coefficients of the model are identified using physics informed regression from real life towing experiment data. Using Shapley value analysis, the model is shown to be able to significantly explain the measured forces and moments on the towing model. Lastly, the banking coefficients of the model are shown to be non-dimensional, and is therefore invariant to the scale of the vessel as long as the shape is preserved.
\section{Problem Formulation}
The aim of this study is to derive a closed-form model for bank effects that can be used to extend classic manoeuvring models \cite{Fossen2011HandbookControl}
\begin{equation}
    \mathbf{M}\frac{d}{dt}\bm{\nu}(t) = \bm{f}(\bm{\nu}(t),\bm{u}(t)) + \bm{d_{bank}}(\bm{\eta}(t),\bm{\nu}(t),E).
\end{equation}
The modelling of the vessel hydrodynamics is limited to 3 degrees of freedom (DOF),  namely surge, sway, and yaw, since large vessels are mostly stationary in the heave direction. The states of the vessel is the position and orientation $\bm{\eta} = (x,y,\psi) \in \R^3$ in the north-east-down (NED) frame and the velocities $\bm{\nu} = (u,v,r) \in \R^3$ in the body frame. Furthermore, there are a number of control inputs to the vessel $\bm{u}\in\R^{N_u}$. The model comprises of two vector fields, where the mapping $\bm{f}: \R^3 \times \R^{N_u} \rightarrow \R^3$ accounts for the nominal open-water manoeuvring model, and $\bm{d_{bank}}: \R^3 \times \R^3 \times \mathcal{E} \rightarrow \R^3$  models the banking effect perturbation, where the environment $E = (W,D) \in \mathcal{E} \subseteq \R^2_{\geq0}$ comprises the width and depth of the canal.

\subsection{Data Collection and Partitioning}
\begin{figure}[ht]
    \centering
    \includegraphics[width=\linewidth]{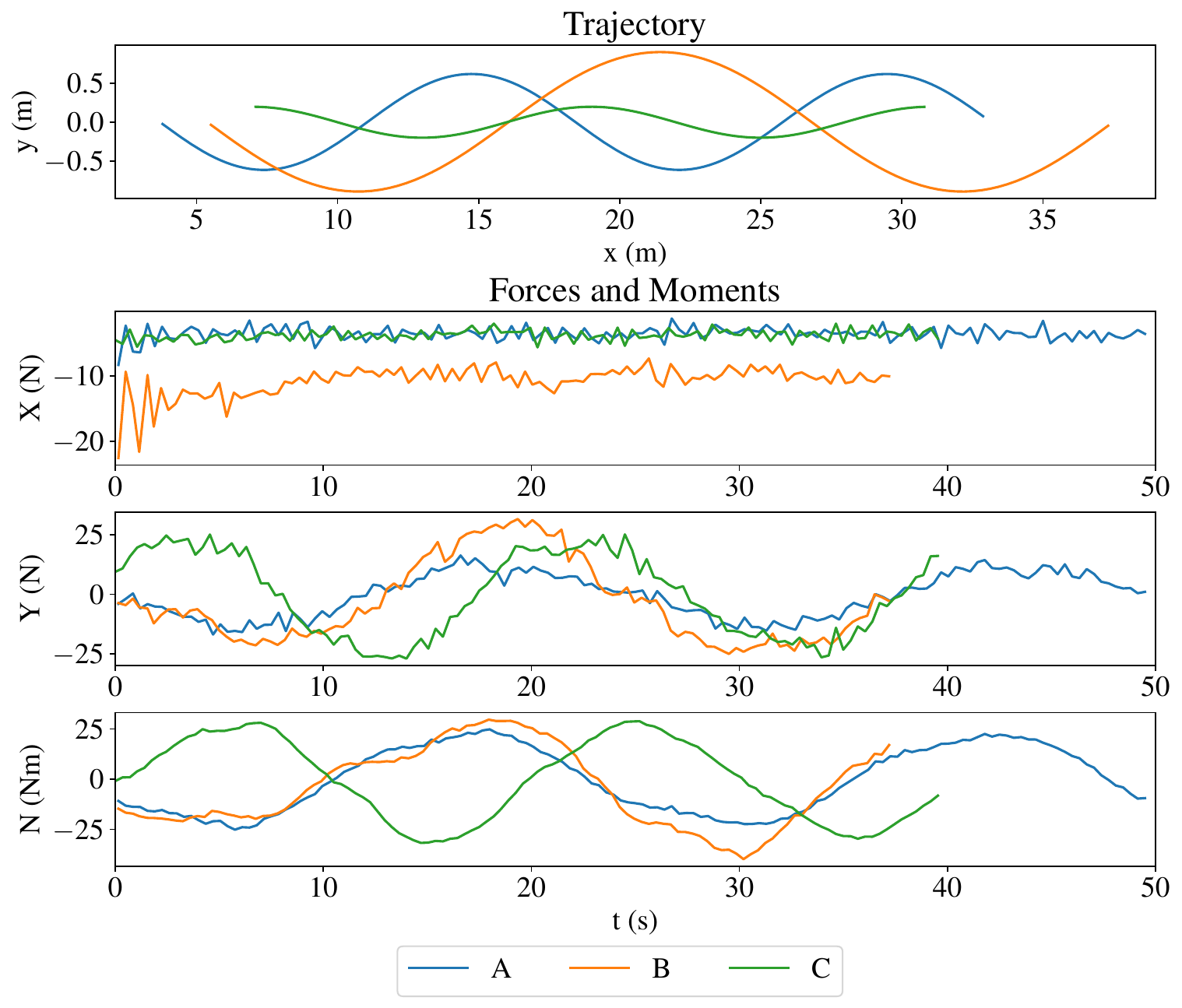}
    \caption{Model ship trajectories, forces, and moments for test scenario A, B and C from \cite{Eloot2016RunningPurposes}.}
    \label{fig:traj_forces}
\end{figure}

To verify the modelling approach, the coefficients for a towing model are determined from experimental data. The towing model is a 1:89.11 scale version of the Duisburg Test Case (DTC) vessel, which is a 14000 TEU Post-Panamax container ship model \cite{ElMoctar2012DuisburgBenchmarking}. The data used in this paper is from \cite{Eloot2016RunningPurposes} where harmonic yaw and sway captive tests were performed in a towing tank. Namely, two tests with harmonic yaw angle and zero sway (test A, B), and a test with harmonic sway and zero rate of turn (test C) were used, see \Cref{fig:traj_forces}. The forces and moments acting on the towing model are measured using a force-torque sensor mounted between the model and the towing rig.

The dataset must be split into a training and validation dataset. The data is split by randomly sampling 80 \% of the data for training and 20\% for validation, to ensure that both the training and validation data is sufficiently excited and to remove the temporal aspect of the data which will not be used in the modelling and estimation. 
\section{Modelling}
The measured induced surge and sway forces $X(t)$, $Y(t)$, and yaw moment $N(t)$ is due to the following three effects:
\begin{enumerate}
    \item Measured force and moment due to the vessel being accelerated
    \item Damping due to the viscosity of the water
    \item Banking due to Bernoulli's principle\cite{Kobylinski2014BankVessels}
\end{enumerate}
Banking effects comprise two phenomena:
\paragraph{\textbf{Bank Suction}}
The water on either side of a vessel in a canal must accelerate to pass through the confined space, reducing the pressure near the vessel's side closest to the bank, pulling it toward the bank.
\paragraph{\textbf{Bank Cushioning}}
Water accumulates along the bow, increasing pressure on the near-bank side of the bow, inducing a yaw moment that pushes the bow away from the bank. The banking only affects the sway force and yaw moment. Therefore, the surge and sway force, and yaw moment acting on the vessel are respectively modelled as:
\begin{subequations}
    \begin{equation}
        X = X_{acc} + X_{damp},
    \end{equation}
    \begin{equation}
        Y = Y_{acc} + Y_{damp} + Y_{bank},
    \end{equation}
    \begin{equation}
        N = N_{acc} + N_{damp} + N_{bank}.
    \end{equation}
\end{subequations}
\subsection{Force and Moment due to Towing Rig Acceleration}
Since the vessel is constrained to move along a predefined trajectory by being pulled on a towing rig, the force and moment sensor mounted between the vessel and the rig will register a force and moment due to the vessel resisting being accelerated, which can be modelled as:
\begin{subequations}\label{eq:mass_inertia}
    \begin{equation}\label{eq:mass_inertia_x}
        X_{acc} = -(m+m_{\dot{u}})\dot{u} = -a_{\dot{u}}\dot{u},
    \end{equation}
    \begin{equation}\label{eq:mass_inertia_y}
        Y_{acc} = -(m+m_{\dot{v}})\dot{v} - (m x_G + m_{\dot{v}\dot{r}})\dot{r} = -b_{\dot{v}}\dot{v} - b_{\dot{r}}\dot{r},
    \end{equation}
    \begin{equation}\label{eq:mass_inertia_z}
        N_{acc} = - (m x_G + m_{\dot{v}\dot{r}})\dot{v} + (I_{z} + I_{\dot{r}})\dot{r} = -c_{\dot{v}}\dot{v} - c_{\dot{r}}\dot{r},
    \end{equation}
\end{subequations}
where $m \in \R_{\geq 0}$ is the mass of the vessel, $I_{z} \in \R_{\geq 0}$ is the inertia of the vessel, $x_G \in \R$ is the distance from the towing rig to the centre of gravity along the x-axis of the vessel, and $m_{\dot{u}},\ m_{\dot{v}},\ m_{\dot{v}\dot{r}},\ I_{\dot{r}} \in R_{\geq 0}$ is the added mass and inertia of the water in the surge, sway and yaw directions. Since regression will be used to identify the terms, they are absorbed into the coefficients $a_{\dot{u}}$, $b_{\dot{v}}$, $b_{\dot{r}}$, $c_{\dot{v}}$, and $c_{\dot{r}}$. Lastly, from \eqref{eq:mass_inertia} it is clear to see that $a_{\dot{u}}, b_{\dot{v}}, c_{\dot{r}} \geq 0$ and $b_{\dot{r}} = c_{\dot{v}}$.
\subsection{Damping Force and Moment}
When the vessel moves through water, it will be subject to damping forces that are dependent on its velocities\cite{Fossen2011HandbookControl}, which can be modelled as:
\begin{subequations}
    \begin{equation}
        X_{damp} = - a_{u}u - a_{|u|u}|u|u,
    \end{equation}
    \begin{equation}
        \begin{split}
            Y_{damp} = -b_{v}v - b_{r}r - b_{|v|v}|v|v - b_{|r|r}|r|r,
        \end{split}
    \end{equation}
    \begin{equation}
        \begin{split}
            N_{damp} = -c_{v}v - c_{r}r - c_{|v|v}|v|v - c_{|r|r}|r|r.
        \end{split}
    \end{equation}
\end{subequations}
For the system to be stable, the following damping coefficients must be positive
\begin{equation}
    a_u, \ a_{|u|u}, \ b_v, \ b_{|v|v}, \ c_r, \ c_{|r|r} \geq 0,
\end{equation}
and the remaining damping coefficients are either positive or negative depending on whether the centre of pressure is in front of or behind the centre of mass. Normally, there would also be damping cross terms between sway and yaw. However, since they are excited independently in the data it is not possible to estimate these.
\subsection{Banking Force and Moment}
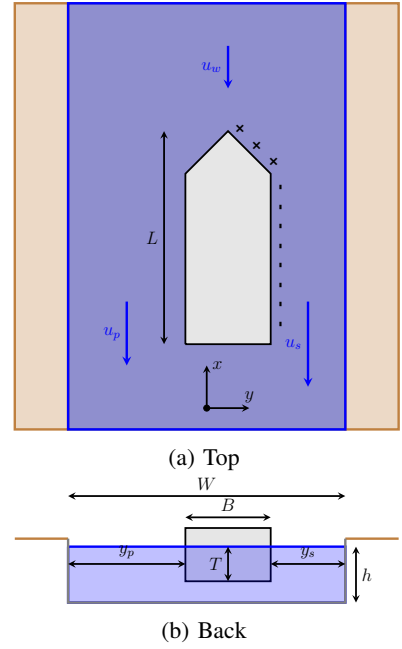
\begin{figure}[t!]
    \centering
    \begin{subfigure}[t]{\linewidth}
        \centering
        \resizebox{0.6\linewidth}{!}{
        \begin{tikzpicture}[font=\large]
            \draw[color=gray,fill=gray!50,ultra thick] (1.25,0) rectangle (7.75,10);
            \draw[color=brown,fill=brown!30,ultra thick] (0,0) rectangle (1.25,10);
            \draw[color=brown,fill=brown!30,ultra thick] (7.75,0) rectangle (9,10);
            \draw[color=blue,fill=blue,fill opacity=0.25,ultra thick] (1.25,0) rectangle (7.75,10);
        
            \draw[color=black,fill=gray!20, very thick] (4,2) -- (4,6) -- (5,7) -- (6,6) node[midway,above,sloped] {\textbf{+ \ + \ +}} -- (6,2) node[midway,sloped,above,align=right] {\textbf{- \ - \ - \ - \ - \ - \ - \ - }} -- (4,2);
        
            \draw[color=blue,-stealth, ultra thick] (5,9) -- (5,8) node[midway,left] {$u_w$};
        
            \draw[color=blue,-stealth, ultra thick] (2.625,3) -- (2.625,1.5) node[midway,left] {$u_p$};
        
            \draw[color=blue,-stealth, ultra thick] (6.875,3) -- (6.875,1.0) node[midway,left] {$u_s$};
        
            \draw[color=black,stealth-stealth,very thick] (3.5,2) -- (3.5,7) node[midway,left] {$L$};
    
            \draw[color=black,-stealth,very thick] (4.5,0.5) -- (5.5,0.5) node[pos=1.0,above] {$y$};
    
            \draw[color=black,-stealth,very thick] (4.5,0.5) -- (4.5,1.5) node[pos=1.0,right] {$x$};
    
            \filldraw[black] (4.5,0.5) circle (2pt);
            
        \end{tikzpicture}}
        \caption{Top}
        \label{fig:banking_top}
    \end{subfigure}
    \begin{subfigure}[t]{\linewidth}
        \centering
        \resizebox{0.6\linewidth}{!}{
        \begin{tikzpicture}[font=\large]
            \draw[color=black,fill=gray!20,very thick] (4,-1.0) rectangle (6,0.25);
            \draw[color=blue,fill=blue,fill opacity=0.25,ultra thick] (1.25,-0.1875) -- (7.75,-0.1875) -- (7.75,-1.5) -- (1.25,-1.5) -- (1.25,-0.1875);
            
            \draw[color=brown,ultra thick] (0,0) -- (1.25,0);
            \draw[color=brown,ultra thick] (7.75,0) -- (9,0); 
            \draw[color=gray,ultra thick] (1.25,0) -- (1.25,-1.5) -- (7.75,-1.5) -- (7.75,0);
            \draw[color=black,very thick,stealth-stealth] (5.0,-0.1875) -- (5.0,-1.0) node[midway,left] {$T$};
            
            \draw[color=black,very thick,stealth-stealth] (1.25,-0.59375) -- (4,-0.59375) node[above,midway] {$y_{p}$};
            
            \draw[color=black,very thick,stealth-stealth] (6,-0.59375) -- (7.75,-0.59375) 
            node[above,midway] {$y_{s}$};
            
            \draw[color=black,very thick,stealth-stealth] (8,-0.1875) -- (8,-1.5) node[midway,right] {$h$};
            
            \draw[color=black,very thick,stealth-stealth] (4,0.5) -- (6,0.5) node[midway,above] {$B$};
            \draw[color=black,very thick,stealth-stealth] (1.25,1) -- (7.75,1) node[midway,above] {$W$};
         \end{tikzpicture}}
        \caption{Back}
        \label{fig:banking_back}
    \end{subfigure}
    \caption{Banking scenario top and back view.}
    \label{fig:banking}
\end{figure}

\begin{figure}[ht]
    \centering
    \resizebox{0.6\linewidth}{!}{
    \begin{tikzpicture}
        \begin{axis}[domain=-3.0:3.0, 
                     samples=400, 
                     axis lines=middle,
                     axis lines=middle,
                     xlabel={\Large $y$ (m)},
                     ylabel={\Large $\delta$ ()},
                     xlabel style={right},
                     ylabel style={above}]

            \filldraw[fill=green, opacity=0.1, very thick] (-0.9068,-4.5) rectangle (0.9068,4.5);
            \draw[color=green!40,very thick] (-0.9068,-4.5) -- (-0.9068,4.5);
            \draw[color=green!40,very thick] (0.9068,-4.5) -- (0.9068,4.5);
       
            \addplot[blue, very thick] {(7/2 - x)^2/(7/2-0.572/2 - x)^2 - (7/2 + x)^2/(7/2-0.572/2 + x)^2};
            \addplot[red, very thick] {((7/2 - x)^2/(7/2-0.572/2 - x)^2 - (7/2 + x)^2/(7/2-0.572/2 + x)^2)*(0.5-0.5*tanh(5*(abs(x)-2.5)))};

        \end{axis}
        
    \end{tikzpicture}}
    \caption{$\delta$ \eqref{eq:Delta_1} as a function of vessel deviation from centre of canal $y$ with parameters taken from the experiments $W=7$ and $B=0.572$ (blue), and with boundary layer effect (red). Green area is the scope of the $y$ experiment data.}
    \label{fig:Delta_fun}
\end{figure}
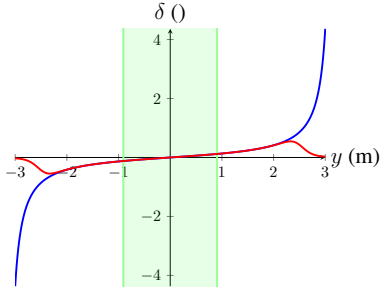

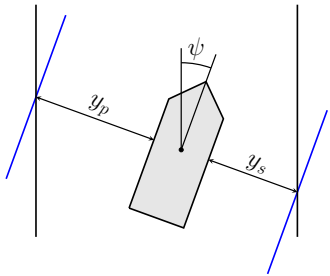
\begin{figure}[ht]
    \centering
    \resizebox{0.5\linewidth}{!}{
    \begin{tikzpicture}
        \draw[color=black,fill=gray!20, ultra thick, rotate=-20] (-1,-2.5) -- (-1,1.5) -- (0,2.5) -- (1,1.5) -- (1,-2.5) -- (-1,-2.5);
        \draw[color=black,fill=black] (0,0) circle (2pt);
        
        \draw[color=black,rotate=-20,very thick,stealth-stealth] (1,0) -- (4.256711,0) node[rotate=-20,midway,above] {\Huge $y_{s}$};

        
        \draw[color=black,rotate=-20,very thick,stealth-stealth] (-1,0) -- (-5.320889,0) node[rotate=-20,midway,above] {\Huge $y_{p}$};

        \draw[color=blue,rotate=-20,ultra thick] (-5.320889,-3) -- (-5.320889,3);

        \draw[color=blue,rotate=-20,ultra thick] (4.256711,-3) -- (4.256711,3);


        \draw[color=black,ultra thick] (-5,-3) -- (-5,5);
        \draw[color=black,ultra thick] (4,-3) -- (4,5);
        
        \draw[color=black,very thick] (0,0) -- (0,3.5);
        \draw[color=black,very thick,rotate=-20] (0,0) -- (0,3.5);
        \draw[color=black,very thick,rotate=90] (3,0) arc (0:-20:3) node[midway,above] {\Huge $\psi$};
    \end{tikzpicture}}
    \caption{Starboard and port side midship hull clearance with non-zero heading $\psi$. The blue line indicates the equivalent aligned canal.}
    \label{fig:clearance_w_heading}
\end{figure}

\begin{figure}[ht]
    \centering
    \includegraphics[width=.6\linewidth]{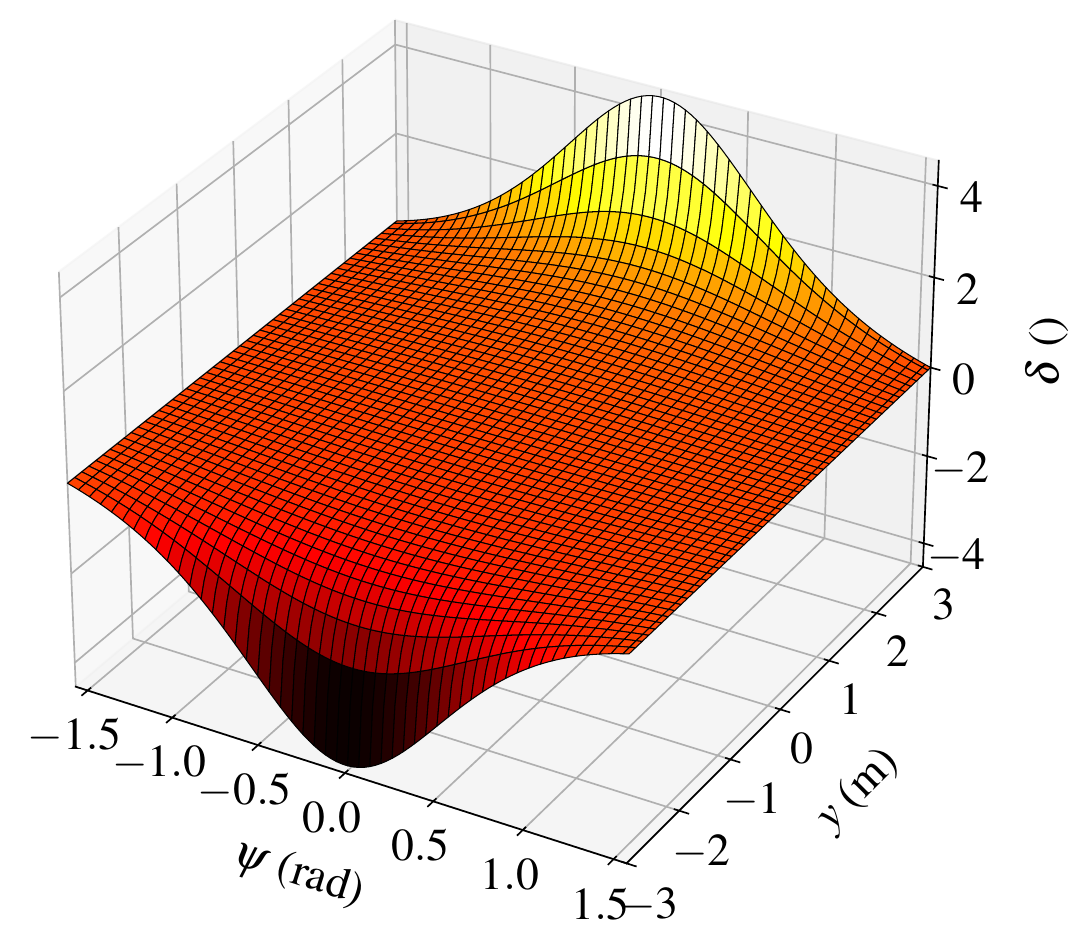}
    \caption{$\delta(y,\psi)$ with varying transverse positions and headings.}
    \label{fig:delta_surf}
\end{figure}

It is initially assumed that the vessel is aligned with the side of the banks. To model the bank effects, the pressure along the hull must be known. Bernoulli's principle relates the velocity of a fluid flow $\upsilon$ to pressure $P$ according to
\begin{equation}
    P + \frac{1}{2}\rho \upsilon^2 + \rho g h + c = 0,
\end{equation}
where $\rho$ is the density of the fluid, $g$ the gravitational constant, $h$ the height of the fluid above a reference plane, and $c$ is a constant. For simplicity, the side of the vessel is assumed to be completely flat. Therefore, the pressure difference along the side of the vessel can be derived as
\begin{equation}\label{eq:pressure_diff_1}
    \begin{aligned}
        \Delta P = P_p - P_s = \frac{1}{2}\rho(u_s^2 - u_p^2),
    \end{aligned}
\end{equation}
where $u_s$ and $u_p$ are the average speed of the water relative to the vessel, passing on the starboard and port side respectively, see \cref{fig:banking_top}. It is assumed that the water moving toward the vessel is funnelled evenly around it. Therefore, due to the continuity principle, the average speed of the water passing on either side of the vessel relative to the vessel can be found as
\begin{equation}
    u_s = \frac{y_{s} + B/2}{y_{s}}u_w, \quad u_p = \frac{y_{p} + B/2}{y_{p}}u_w,
\end{equation}
where $B$ is the beam of the vessel, and $y_{s3}$ and $y_{p3}$ are the starboard and port side hull clearances respectively, see \cref{fig:banking}. Inserting this into \eqref{eq:pressure_diff_1} gets
\begin{equation}
    \Delta P = \frac{1}{2}\rho\delta(y) u_w^2,
\end{equation}
where
\begin{equation}\label{eq:Delta_1}
    \delta(y) = \frac{(y_{s} + B/2)^2}{y_{s}^2} - \frac{(y_{p} + B/2)^2}{y_{p}^2}.
\end{equation}
As can be seen in \cref{fig:Delta_fun}, $\delta$ remains almost linear in the domain around $y=0$, and grows unbounded as it approaches the boundary $\pm (W/2 - B/2)$ of its domain. According to this model, the velocity of the water on the side of the vessel closest to the bank will grow to infinity as the bank clearance goes towards zero. Due to the viscosity of the water, after a certain distance the amount of water passing between the ship and the bank will go to zero because of the boundary layer effect. This results in the model seen on \cref{fig:Delta_fun}.
The exact distance at which the boundary layer effect occurs is unknown. However, it is assumed that at the point where the boundary layer effect becomes active, grounding is imminent and that there is no point in trying to model the banking effect in that domain. Therefore, the model in \eqref{eq:Delta_1} is assumed to be valid in the region around the centre of the canal where the traffic normally is. Assuming a rectangular side profile, the bank suction force being applied to the vessel can then be found as
\begin{equation}\label{eq:bank_sway_w_assum}
    \begin{aligned}
        Y_{bank} &= b_{bank} C_BL(T_0 + z)\Delta P \\
        &= b_{bank} \frac{1}{2} C_B \rho L(T_0 + z) \delta u_w^2,
    \end{aligned}
\end{equation}
where the pressure difference $\Delta P$ is multiplied by the length of the vessel $L$, the nominal draft of the vessel $T_0$ plus the heave $z$, the block coefficient $C_B \in [0,1$], and $b_{u_w^2}$ a parameter that depends on specifics of the vessel and the canal. The block coefficient is the ratio between the displacement of the vessel and the submerged volume of its minimum bounding box, capturing how close the shape is to the idealised case. This is multiplied into the banking force to lift the approximation of the vessel being a box. The bank cushioning moment can be calculated similarly
\begin{equation}\label{eq:bank_yaw_w_assum}
    N_{bank} = c_{bank} \frac{1}{2} C_B \rho L^2(T_0 + z)\delta u_w^2,
\end{equation}
where $c_{u_w^2}$ is a parameter that depends on specifics of the vessel and the canal, and the pressure difference $\Delta P$ is multiplied by the square of the ship length. Finally, the assumption of vessel-canal alignment needs to be lifted. When the vessel rotates, the distance from the midship hull to the bank, as well as the distance from the bow to the bank increases according to the heading $\psi$, see \cref{fig:clearance_w_heading}.
The midship hull clearances can be calculated according to
\begin{subequations}\label{eq:clearance_w_heading}
    \begin{equation}
        y_{s} = (W/2 - y)\sec|\psi| - B/2,
    \end{equation}    
    \begin{equation}
        y_{p} = (W/2 + y)\sec|\psi| - B/2.
    \end{equation}  
\end{subequations}
Assuming that the canal with the non-zero heading can be modelled as an equivalent aligned canal, see \cref{fig:clearance_w_heading}, and inserting \eqref{eq:clearance_w_heading} into \eqref{eq:Delta_1} gets
\begin{equation}
    \begin{aligned}
        \delta(y,\psi) =& \frac{((W/2 - y)\sec|\psi|)^2}{((W/2 - y)\sec|\psi| - B/2)^2} - \\ & \frac{((W/2 + y)\sec|\psi|)^2}{((W/2 + y)\sec|\psi| - B/2)^2}. 
    \end{aligned}
\end{equation}
which can be seen plotted in \cref{fig:delta_surf}.
As before, $\delta$ grows unbounded as $y$ approaches the boundary of its domain for $\psi=0$. However, as $|\psi|$ approaches $\pi/2$, the width of the equivalent canal goes to infinity and therefore $\delta$ decays to zero. This fits well with the expected behaviour, that when the vessel is oriented transverse in the canal there will be no banking.
\section{Parameter Identification}
Having derived candidate models for the sway force and yaw moment, the regressors $\bm{a}$, $\bm{b}$, and $\bm{c}$ must be determined from the data. To that end, each term of the candidate functions are stacked together into three candidate regression matrices
\begin{subequations}
    \begin{gather}
        \begin{split}
            \mathbf{\Theta_X} = [ \ -\bm{\dot{u}} \quad -\bm{u} \quad -|\bm{u}|\bm{u} \ ] \in \R^{M \times 3}
        \end{split} \\
        \begin{split}
            \mathbf{\Theta_Y} = [ &
                \ -\bm{\dot{v}} \quad -\bm{\dot{r}} \quad -\bm{v} \quad -\bm{r} \quad -\bm{|v|v} \quad -\bm{|r|r} \\ & \frac{1}{2}C_B \rho L(T_0+\bm{z})\bm{\delta}(\bm{y},\bm{\psi}) \bm{ u_w^2} \ ] \in \R^{M\times7},
        \end{split} \\
        \begin{split}
            \mathbf{\Theta_N} = [ &
                \ -\bm{\dot{v}} \quad -\bm{\dot{r}} \quad -\bm{v} \quad -\bm{r} \quad -\bm{|v|v} \quad -\bm{|r|r} \\ & -\frac{1}{2}C_B \rho L^2(T_0+\bm{z})\bm{\delta}(\bm{y},\bm{\psi}) \bm{ u_w^2} \ ] \in \R^{M\times7},
        \end{split} 
    \end{gather}
\end{subequations}
where the bold symbol denotes that it is the data from the experiments gathered into column vectors, and the operations are performed element wise. 
The expected measured forces and moment on the model can then be found as
\begin{equation}\label{eq:exp_force}
    \bm{\hat{X}} = \mathbf{\Theta_X} \bm{a}, \quad 
    \bm{\hat{Y}} = \mathbf{\Theta_Y} \bm{b}, \quad
    \bm{\hat{N}} = \mathbf{\Theta_N} \bm{c},
\end{equation}
where
\begin{subequations}
    \begin{gather}
        \bm{a} = (m_{u}, a_{u}, a_{|u|u}), \\
        \bm{b} = (m_{v}, b_{\dot{r}}, b_v, b_r, b_{|v|v}, b_{|r|r}, b_{bank}), \\
        \bm{c} = (c_{\dot{v}}, I_{zz}, c_v, c_r, c_{|v|v}, c_{|r|r}, c_{bank}).
    \end{gather}
\end{subequations}
The goal is then to determine the optimal regressors from the data. This can be done using optimisation, where the regressors that minimise the mean squared error (MSE) between the measured and expected force and moment is determined. The optimal regressors $\bm{a^*}$, $\bm{b^*}$, and $\bm{c^*}$ can then be calculated as the solution to the following optimisation problems
\begin{equation}\label{eq:opt_prob}
    \begin{aligned}
        \bm{a^*}, \bm{b^*}, \bm{c^*} = \arg & \min_{\bm{a}, \bm{b}, \bm{c}} \frac{1}{K} \left( \right. ||\bm{X} - \mathbf{\Theta_X}\bm{a}||_2^2 + \\
        &\qquad\quad\ \ \ \ ||\bm{Y} - \mathbf{\Theta_Y}\bm{b}||_2^2 + \\
        &\qquad\quad\ \ \ \ ||\bm{N} - \mathbf{\Theta_N}\bm{c}||_2^2 \left. \right ) \\
        s.t. & \\
        & b_{\dot{r}} = c_{\dot{v}}, \\
        & a_u,\ a_{|u|u},\ b_v,\ b_{|v|v},\ c_r,\ c_{|r|r} \geq 0.
    \end{aligned}
\end{equation}
Before estimating the optimal coefficients, it is prudent to investigate correlations between the candidate functions and whether the problem is ill-conditioned or not.
\begin{figure}[b!]
    \centering
    \includegraphics[width=\linewidth]{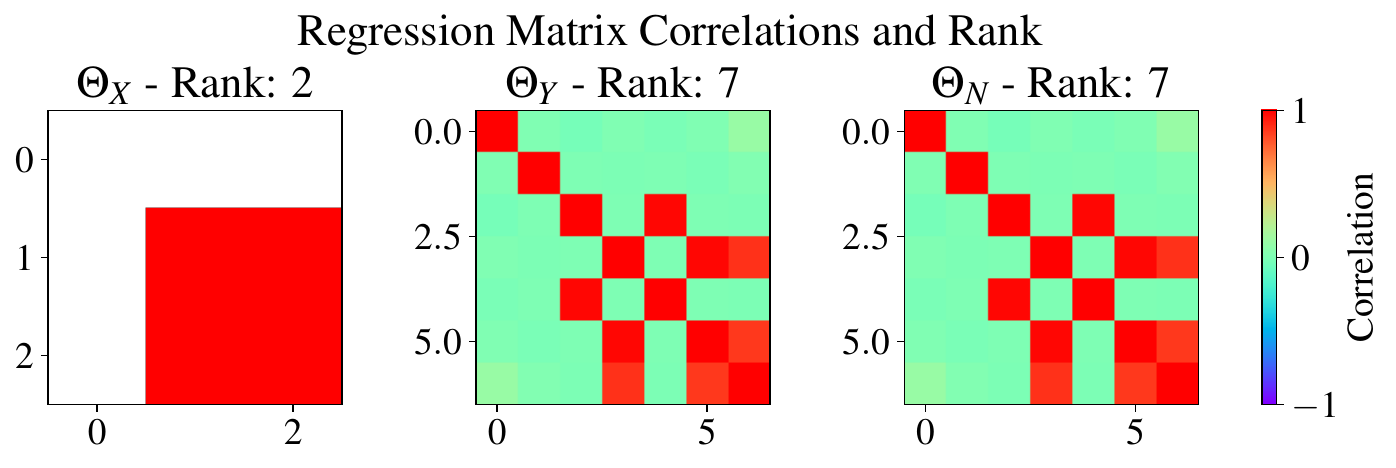}
    \caption{Correlations between candidate functions for surge, sway, and yaw, and rank of the regression matrices.}
    \label{fig:corr}
\end{figure}
As can be seen in \cref{fig:corr}, there are some correlations between the candidate functions, but the regression matrixes $\bm{\Theta_Y}$ and $\bm{\Theta_N}$ have full rank. The regression matrix $\bm{\Theta_X}$ does not have full rank. This is due to the fact that the surge velocity is constant, meaning that the surge acceleration is zero. This results in the regression coefficient $m_{u}$ being unidentifiable.
\section{Results}
\begin{figure}[ht]
    \centering
    \includegraphics[width=\linewidth]{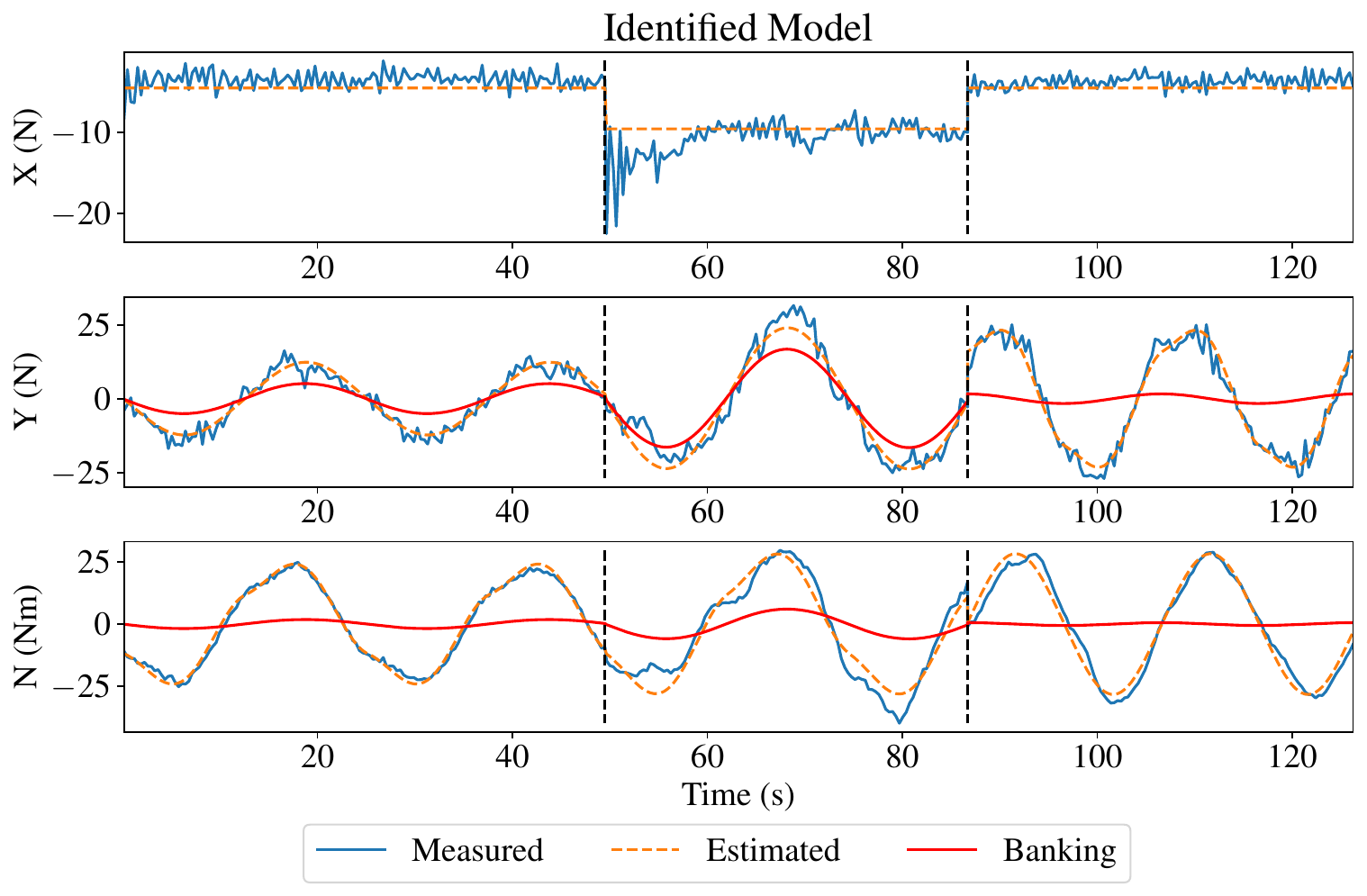}
    \caption{Measured forces and moments, estimated forces and moments from the identified model, and estimated banking force and moment. The vertical black dashed lines indicate the different datasets.}
    \label{fig:result}
\end{figure}

Solving the optimisation problem in \eqref{eq:opt_prob} using the experimental data results in the following optimal regressors
\begin{subequations}
    \begin{gather}
        \bm{a^*} = \left (\ 0, 0, 12.6 \ \right ), \\
            \bm{b^*} = \left( \ 733, -56.1, 100, 118, 3298, -161, 1.07 \ \right), \\
            \bm{c^*} = \left( \ -56.1, 712, 414, 84.9, 589, 3346, 0.13 \ \right).
    \end{gather}
\end{subequations}
To validate the significance of the proposed banking model a Shapley value test is performed. In the Shapley value test, the marginal contributions of each column of the validation regression matrixes are determined. All the values are normalised by the sum of the absolute Shapley values. Therefore, a value of $0$ indicates that the column is insignificant and a value of $1$ means that it is the only term that is significant.
\begin{table}[h]
    \centering
    \caption{Shapley Value Test Results}
    \label{tab:significance_test}
    \begin{subtable}[t]{0.32\linewidth}
        \centering
        \begin{tabular}{c|c}
                  & \textbf{Shapley} \\
            \hline
            \hline
            $a_{\dot{u}}$ & 0  \\
            $a_u$ & 0  \\ 
            $a_{|u|u}$ & 1.0  \\
        \end{tabular}
    \end{subtable}
    \hspace{\fill}
    \begin{subtable}[t]{0.32\linewidth}
        \centering
        \begin{tabular}{c|c}
                  & \textbf{Shapley} \\
            \hline
            \hline
            $b_{\dot{v}}$ & 0.239 \\ 
            $b_{\dot{r}}$ & 0.002 \\ 
            $b_{v}$ & 0.087 \\ 
            $b_{r}$ & 0.243 \\ 
            $b_{|v|v}$ & 0.132 \\ 
            $b_{|r|r}$ & 0.015 \\ 
            $b_{bank}$ & 0.283 \\ 
        \end{tabular}
    \end{subtable}
    \hspace{\fill}
    \begin{subtable}[t]{0.32\linewidth}
        \centering
        \begin{tabular}{c|c}
                  & \textbf{Shapley} \\
            \hline
            \hline
            $c_{\dot{v}}$ & 0.040 \\ 
            $c_{\dot{r}}$ & 0.255 \\ 
            $c_{v}$ & 0.235 \\ 
            $c_{r}$ & 0.128 \\ 
            $c_{|v|v}$ & 0.026 \\ 
            $c_{|r|r}$ & 0.248 \\ 
            $c_{bank}$ & 0.069 \\ 
        \end{tabular}
    \end{subtable}
\end{table}
As can be seen in \Cref{tab:significance_test}, both the banking terms for the sway force and yaw moment are significant. The Shapley value for the banking sway force $b_{bank}$ is approximately $28\%$, and for the yaw moment $c_{bank}$ it is $7\%$. Lastly, the total estimated forces and moments, and the banking force and moment can be seen in \cref{fig:result}. As can be seen, the model is able to recreate the measured forces and moments.

\section{Nondimensional Model Coefficients}
Having derived a hydrodynamic manoeuvring model for the towing model, the hydrodynamic coefficients must be non-dimensionalised such that they can be applied to full scale vessels. To non-dimensionalise the coefficients, the units of each coefficient is found and substituted with the ships parameters according to a nondimensionalisation system of choice. For brevity, the procedure for nondimensionalising the mass-inertia and damping coefficients is omitted, but can be found in, e.g., \cite{Fossen2011HandbookControl}. However, it will be demonstrated that the coefficients of the banking model need not be nondimensionalised. For the banking sway force, the units of the different factors can be written out as
\begin{equation}
    \underbrace{Y_{bank}}_{\frac{kg m}{s^2}} = b_{bank}\frac{1}{2}\underbrace{C_B}_{1}\underbrace{\rho}_{\frac{kg}{m^3}} \underbrace{L(T_0 + z)}_{m^2}\underbrace{\delta}_{1}\underbrace{u_w^2}_{\frac{m^2}{s^2}}.
\end{equation}
Multiplying the units together on the RHS of the equation gets $\frac{kg}{m^3} m^2 \frac{m^2}{s^2} = \frac{kg m}{s^2}$, thereby making the coefficient $b_{bank}$ nondimensional. The same can be applied to the banking yaw moment
\begin{equation}
    \underbrace{N_{bank}}_{\frac{kg m^2}{s^2}} = -c_{bank}\frac{1}{2}\underbrace{C_B}_{1}\underbrace{\rho}_{\frac{kg}{m^3}} \underbrace{L^2(T_0 + z)}_{m^3}\underbrace{\delta}_{1}\underbrace{u_w^2}_{\frac{m^2}{s^2}}.
\end{equation}
where the units multiplied together on the RHS of the equation gets $\frac{kg}{m^3} m^3 \frac{m^2}{s^2} = \frac{kg m^2}{s^2}$, thereby making the coefficient $c_{bank}$ nondimensional. Therefore, it is assumed that the banking coefficients are applicable to all scales of ships and banks, as long as the relative geometries are preserved.

\section{Simulation}
\begin{figure}[ht]
    \centering
    \includegraphics[width=\linewidth]{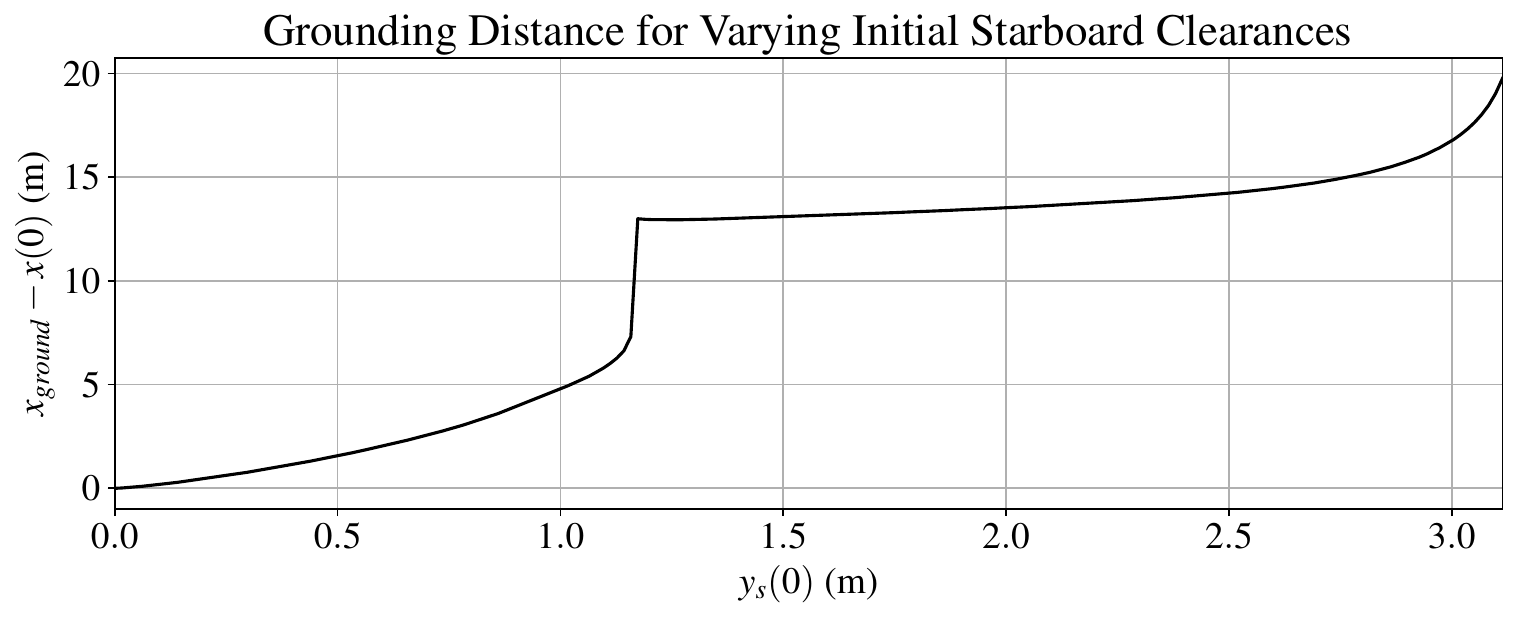}
    \caption{Distance in the direction of the canal at which the vessel grounds given an initial starboard clearance $y_s(0)$ and a velocity of $u(0)=1 \ m s^{-1}$.}
    \label{fig:ground_dist}
\end{figure}

The identified model for the scaled vessel is simulated with varying initial positions $y(0)$. The considered scenario emulates a vessel moving in a canal with a width of $7$ m and with initial surge velocity of $u(0) = 1 \ ms^{-1}$. The input surge force is kept constant at $X_{in} = 12.6 \ kg m s^{-2}$ to ensure that the vessel maintains a constant surge velocity. The initial transverse position $y_0$ ranges from $0.1 \ m$ to $2.5 \ m$. The remaining states are initialised to zero.
\begin{figure}[h]
    \centering
    \includegraphics[width=\linewidth]{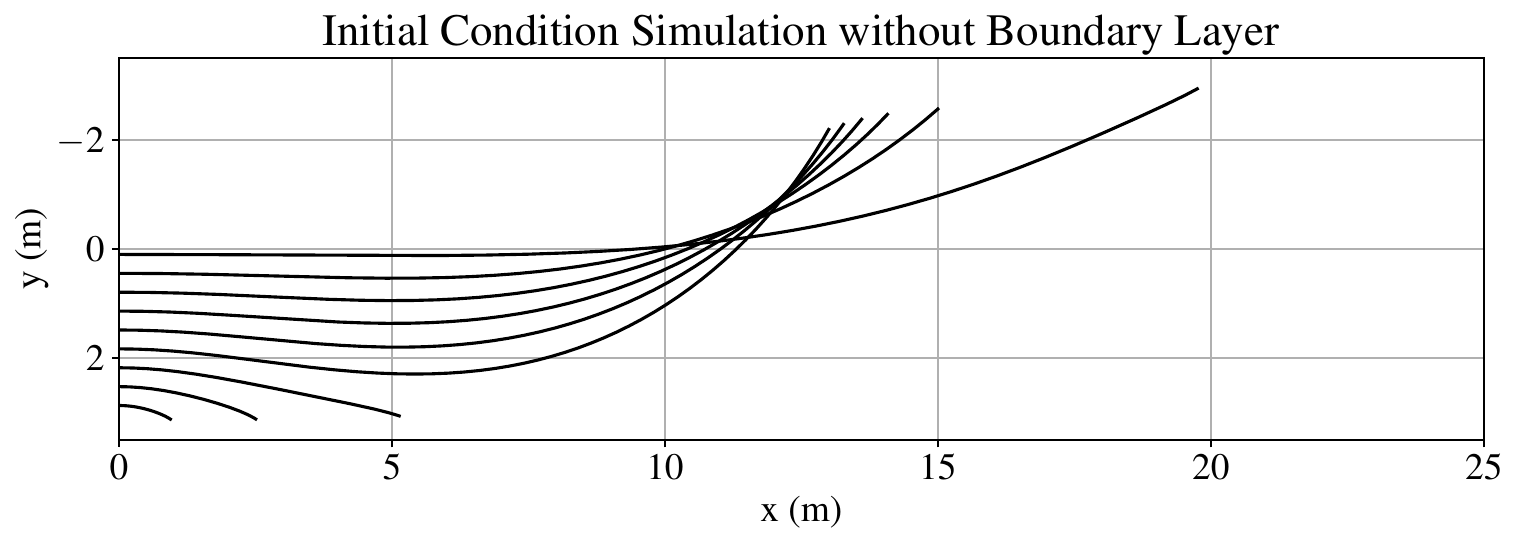}
    \caption{Simulation of towing model with varying initial transverse positions $y_0$.}
    \label{fig:sim}
\end{figure}
As can be seen in \cref{fig:sim}, the banking effect becomes more pronounced as the initial position gets further away from the centre of the canal, leading to the vessel grounding earlier. On \Cref{fig:ground_dist}, the $x$ distance along the canal where the vessel grounds given the initial starboard bank clearance $y_s(0)$ can be seen. When the initial transverse position $y_0$ is approximately greater than $2 \ m$, the vessel grounds on the starboard bank, due to it not turning sufficiently fast. This corresponds to the discontinuity on \Cref{fig:ground_dist} at around $y_{s}(0) \approx 1.2 \ m$. Furthermore, even for a small initial position of $y_0 = 0.1 \ m$, the banking effect leads to grounding. This is due to the system being highly unstable, with the banking effects compounding over time, and with no preventative action, even a small heading angle will inevitably lead to grounding.

\section{Conclusion}
A first-principles closed-form model of banking effects was proposed in this paper that can be used for early warning systems and control design during navigation in confined waters. The coefficients for a towing model were identified based on experimental data using physics-informed regression. A Shapley value test was employed to demonstrate the significance of the marginal contribution of the banking terms to the identified model, which, consequently, validated the proposed model. Finally, it was shown that the coefficients for the proposed model are nondimensional, and can therefore be applied to vessels of any scale, as long as the relative geometries are kept identical. Having derived and identified a closed-form banking model, this can be used for control and detection, such as a bank effect warning system.
\section{Acknowledgement}
The findings in this study are part of the research activities within the ENDYPOS-M$^2$ project, sponsored by Orients Fond Denmark and supported by DTU Maritime, in collaboration with Force Technology, Vessel Performance Solutions, and ABB A/S. The data used for this study was provided by Flanders Hydraulics Research.

\bibliographystyle{IEEEtran}
\bibliography{references}

\end{document}